\documentclass[english,twocolumn,aps,prl,10pt]{revtex4-2}

\usepackage[T1]{fontenc}
\usepackage[latin1]{inputenc}
\usepackage{graphicx}
\usepackage{amssymb}
\usepackage{amsmath}
\usepackage{bm}

\usepackage{babel}
\makeatother

\begin{document}

\title{Dynamics of a Fermi gas quenched to unitarity}

\author{P. Dyke, A. Hogan, I. Herrera, C. C. N. Kuhn, S. Hoinka, and C. J. Vale$^*$}
\affiliation{Optical Sciences Centre, ARC Centre of Excellence in Future Low-Energy Electronics Technologies, Swinburne University of Technology, Melbourne 3122, Australia. \\
{$^\ast$To whom correspondence should be addressed; E-mail: cvale@swin.edu.au} }

\date{\today}

\begin{abstract}
We present an experimental study of a two component Fermi gas following an interaction quench into the superfluid phase. Starting with a weakly attractive gas in the normal phase, interactions are ramped to unitarity at a range of rates and we measure the subsequent dynamics as the gas approaches equilibrium. Both the formation and condensation of fermion pairs are mapped via measurements of the pair momentum distribution and can take place on very different timescales, depending on the adiabaticity of the quench. The contact parameter is seen to respond very quickly to changes in the interaction strength, indicating that short-range correlations, based on the occupation of high-momentum modes, evolve far more rapidly than the correlations in low-momentum modes necessary for pair condensation.

\end{abstract}


\maketitle

Quantum systems far from thermal equilibrium can display features and behaviours that are not captured by equilibrium models \cite{Polkovnikov11,Eisert15}. Examples such as optically induced superconductivity \cite{Fausti11} highlight the potential for exploiting dynamical phenomena in materials driven out of equilibrium. Atomic quantum gases offer unique advantages for building a quantitative understanding of such phenomena \cite{Robertson09} where the timescales for dynamics are typically much longer than comparable scales in solids. Additionally, many of the relevant parameters can be easily tuned allowing access to behaviors that can be difficult to study in other settings \cite{Gring12,Foster14}.

Since the first observation of atomic gas Bose-Einstein condensates (BECs), condensate formation dynamics have attracted strong interest \cite{Miesner98,Kohl04,Smith12,Davis17}. Quenches have enabled studies of features beyond Bogoliubov theory \cite{Papp08,Lopes17a}, vortex formation \cite{Weiler08}, the contact parameter \cite{Wild12,Makotyn14,Fletcher17} and universal dynamics \cite{Erne18,Eigen18,Prufer18}. In Bose gases, however, quenches to the strongly interacting regime are accompanied by rapid inelastic losses via three-body collisions \cite{Rem13,Fletcher13} meaning experiments are typically limited to short timescales. Two-component Fermi gases, on the other hand, are virtually immune to three-body losses \cite{Petrov04} allowing access to dynamics across a broad range of timescales. An early study using modulated interactions identified the timescale for growth of a strongly interacting Fermi condensate \cite{Zwierlein05}. Vortex formation following a rapid final evaporation across the superfluid transition was recently been investigated and seen to display Kibble-Zurek scaling \cite{Ko19,Liu19}. Temperature quenches, however, are limited to rates of the order a few Hz such that the short-time dynamics upon crossing the phase transition have been inaccessible in previous studies. Faster quenches may be achieved by jumping the interaction strength, for example by driving radio-frequency transitions, and was recently used to observe the Higgs amplitude mode through the BEC-BCS crossover \cite{Behrle18} and condensate oscillations \cite{Harrison20}.

In this letter, we study the dynamics of a two-component Fermi gas following a quench from weak to unitarity-limited interactions. We observe that correlations evolve at vastly different rates, depending on the corresponding length scale. The quench involves crossing the normal to superfluid phase transition and we measure the pair momentum distribution to track the formation of a pair condensate. Pairing is seen to take place on short timescales, governed by local properties of the gas, whereas condensation and equilibration of the momentum distribution can take much longer, depending on the adiabaticity of the quench. Using a combination of Tan relations \cite{Tan08b,Tan08c} we show that the contact parameter, that quantifies short-range pair correlations, builds up very rapidly compared to other processes involving larger length scales.

Our experiments are performed with a cloud of $^6$Li atoms prepared in an equal mixture of the $|F = 1/2, m_F =  1/2\rangle$ and $|F = 3/2, m_F =  -3/2\rangle$ hyperfine states. This mixture features a broad Feshbach resonance at 689.7~G enabling precise control of the $s$-wave collisions \cite{Zurn13}. The atoms are confined in an oblate harmonic potential produced by a combination of optical and magnetic fields. A blue-detuned TEM$_{01}$ mode laser beam (with a separation of 49~$\mu$m between antinodes and a horizontal $1/e^2$ radius of 1.1 mm) provides confinement along $z$ with $\omega_z = 2\pi \times 328(1)$~Hz \cite{Smith05,Rath10,Dyke16}. Radial confinement arises from a residual curvature in the magnetic field produced by the Feshbach coils, leading to a highly harmonic radial potential with $\omega_r = \sqrt{\omega_x \omega_y} \approx 2 \pi \times 22$~Hz, where the asymmetry in the trapping potential is $|\omega_x-\omega_y|/\omega_r \lesssim 0.01$. Note that $\omega_r \propto \sqrt{B}$ so the radial confinement also changes when we tune interactions. After evaporative cooling at a magnetic field of 690~G, where the $s$-wave scattering length diverges, $a \rightarrow \infty$, the magnetic field is ramped linearly to 800~G in 200~ms, where $a \approx -3700 a_0$ and $a_0$ is the Bohr radius \cite{Zurn13}, and held for a further 100~ms to ensure the cloud has reached equilibrium. At this point, we observe no dynamics and the interaction parameter is $1/(k_F^{\mathrm{HO}} a) = -1.89 \pm 0.07$, where $k_F^{\mathrm{HO}}$ is the Fermi wavevector given by the Fermi energy in a harmonic trap is $E_F^{\mathrm{HO}} = ( \hbar k_F^{\mathrm{HO}} )^2 / (2 m) = (3 N)^{1/3} \hbar \bar{\omega}$, $m$ is the mass of a lithium atom, $N$ is the total number of atoms and $\bar{\omega} = (\omega_x \omega_y \omega_z)^{1/3}$ is the geometric mean confinement frequency. This lies in the BCS regime where the gas has a similar density distribution to that of an ideal Fermi gas, with perturbative corrections to the energy \cite{Lee58,Giorgini08}. As such we can estimate the temperature $T/T_F^{\mathrm{HO}}$, where $T_F^{\mathrm{HO}} = E_F^{\mathrm{HO}} / k_B$ and $k_B$ is Boltzmann's constant, by fitting the equation of state for an ideal Fermi gas to the measured 2D in-situ density profile. This yields $T/T_F^{\mathrm{HO}} = 0.06 \pm 0.01$ and the interactions reduce the Fermi radius by $\sim 4\%$ compared to an ideal gas \cite{Giorgini08}. This temperature is well above the superfluid transition temperature $\approx 0.02 \, T_F$ for this interaction strength \cite{Haussman07} and we do not observe any evidence for pair condensation prior to the quench.

\begin{figure}[ht]
	\centering
		\includegraphics[clip, width = 1 \columnwidth]{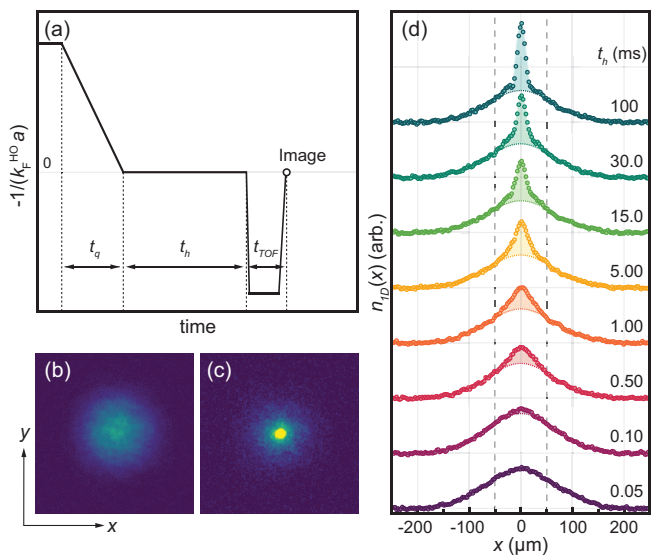}
		\caption{(Color online) (a) Experimental protocol for the quench experiments. A cloud is initially prepared in the weakly interacting BCS regime, with $1/(k_F^{\mathrm{HO}} a) \approx -1.9$, the magnetic field is then ramped to unitarity in a time, $t_q$, and held for $t_h$ before commencing the imaging sequence. (b) and (c) show absorption images of the cloud after $t_q = 50 \; \mu$s and $t_h = 30$ ms, where (b) is an \emph{in situ} image of the trapped cloud and (c) is an absorption image taken after a rapid ramp of the magnetic field deep into the BEC regime, and subsequent time of flight expansion, which reveals the presence of a condensate in the pair-momentum distribution \cite{Regal04}. (d) One-dimensional optical density profiles $n_{\mathrm{1D,TOF}}(x)$ following a 50 $\mu$s quench for a selection of profiles are shown at different hold times, $t_h$. Fine dotted lines show Gaussian fits to the wings of the distributions and the shaded regions indicate the non-Gaussian component $N_{\mathrm{nG}}$. The fitting region excludes data between the dashed vertical lines.}
		\label{fig:fig1}
\end{figure}

After preparing clouds at the $1/(k_F^{\mathrm{HO}} a) = -1.89$ we commence the sequence shown in Fig.~1(a). The magnetic field is ramped linearly to unitarity (690 G) in a time, $t_q$, and the cloud is held for a variable time, $t_h$, before being imaged. Images of the clouds are taken either \emph{in situ} or, following a rapid ramp of the interactions and subsequent time of flight expansion \cite{Regal04}. \emph{In situ} images reveal little of the underlying dynamics, showing only a small amplitude density oscillation for quenches with $t_q \lesssim 2 \pi/\omega_r$, corresponding to the radial monopole mode (a breathing oscillation of the cloud radius in the $x$-$y$ plane) with a frequency of $\sqrt{3} \omega_r/(2 \pi) \approx 38$ Hz \cite{DeRosi15}. Internal dynamics, including pairing and pair-condensation, are nonetheless taking place, but at unitarity, these are virtually undetectable in images of trapped gases. To measure the pair momentum distribution, we suddenly remove the optical ($z$) confinement and jump the magnetic field far to the BEC side of the Feshbach resonance, in less than 100~$\mu$s, which converts weakly bound pairs to tightly bound molecules that preserve their centre of mass momentum. The resultant cloud of weakly interacting molecules is then allowed to expand along $z$ before the field is ramped back to 690~G in 1.5~ms to dissociate the molecules and an absorption image is acquired. The total time from release to imaging corresponds to one quarter of the radial trapping period, $t_{\mathrm{TOF}} = t_r/4 \approx 12$~ms where $t_r = 2 \pi/\omega_r$, in the residual magnetic confinement, such that the spatial distribution reflects the pair momentum distribution prior to release \cite{Murthy14}. Figs.~1(b) and (c) show examples of the \emph{in-situ} and expanded density distributions, respectively, following identical quenches and a hold time of $t_h = 30$~ms. The rapid ramp/time of flight image, (c), shows the characteristic bi-modal distribution of a pair condensate which is not evident in the image of the trapped cloud, (b).

Following a non-adiabatic quench, the kinetic, interaction and potential energies are unbalanced which drives both microscopic and macroscopic dynamics as the cloud evolves towards a new equilibrium. Figure 1(d), shows a selection of 1D density profiles, $n_{\mathrm{1D,TOF}}(x) = \int n_{\mathrm{TOF}}(x,y,z) dy dz$, where $n_{\mathrm{TOF}}(x,y,z)$ is the atom density after the rapid ramp and time of flight expansion, following a $t_q = 50$~$\mu$s quench. Each profile is the average of 3 images taken under the same conditions and provides a representation of the 1D momentum distribution of the pairs. Also shown, are Gaussian fits (fine dotted lines) to these distributions which exclude points between the dashed lines. Immediately following the quench, $t_h$ = 50~$\mu$s, dark purple points in Fig.~1(d), the full cloud is very well described by a Gaussian and is essentially unchanged from its distribution before the quench. For longer hold times, a Gaussian still provides a good fit in the wings but a progressively sharper central peak is seen to develop (shaded regions). We associate the appearance of this peak with the formation of pairs in modes with low centre-of-mass momentum. The peak is initially small but grows rapidly and becomes sharper with increasing hold times. By $t_h \gtrsim 50$~ms this non-Gaussian component takes the shape of an inverted parabola as seen for Fermi condensates at equilibrium \cite{Regal04}. Beyond $\sim 100$~ms hold time, the observed distributions remain stable within our resolution indicating that the cloud has fully equilibrated.

\begin{figure}[ht]
	\centering
		\includegraphics[clip, width = 1 \columnwidth]{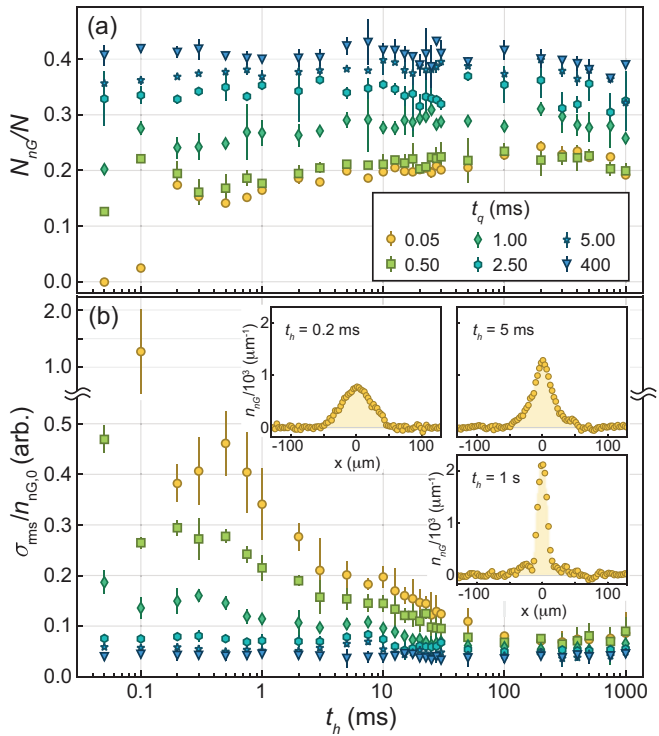}
		\caption{(Color online) (a) Evolution of the non-Gaussian fraction $N_{\mathrm{nG}}/N$ in the pair momentum distribution as a function of hold time, $t_h$, for a selection of different quench durations $t_q$. For quenches longer than a few ms, $N_{\mathrm{nG}}/N$ is approximately equal to $N_{0}/N$ found by fitting a bi-modal distribution (Gaussian plus inverted parabola) to the 1D density profiles. (b) Ratio of the rms width divided by the height of the non-Gaussian component of the momentum distribution. The peak shape is seen to evolve for up to 100 ms following the quench as the occupation of low-momentum modes approaches equilibrium. }
		\label{fig:fig2}
\end{figure}

We have conducted a series of quench experiments and measured the 1D pair momentum distributions, $n_{\mathrm{1D,TOF}}(x)$, for a range of different quench rates. From these we determine the non-Gaussian fraction, $N_{\mathrm{nG}}/N$, given by the area of the central peak above the Gaussian fit, shaded regions in Fig.~1(d), where $N$ is found from the area under the full profile. In Fig.~2(a) we plot $N_{\mathrm{nG}}/N$ as a function of the hold time, for a selection of different quenches, spanning from 50~$\mu$s up to 400~ms. For the fastest quenches, the non-Gaussian fraction appears quite soon after the quench, growing rapidly in the first few 100 $\mu$s, on the order of a few Fermi times. This is much faster than the timescale set by the harmonic trap $\tau_{\mathrm{HO}} = 1/\bar{\omega} \approx 3$ ms and indicates that pair formation is thus governed by local properties of the gas. The Fermi time $\tau_F$ is set by the local Fermi energy $\epsilon_F$, where $\tau_F = \hbar / \epsilon_F$ ($\approx 26$~$\mu$s, at the centre of the cloud), $\epsilon_F = k_B T_F = \hbar^2/(2 m) (3 \pi^2 n)^{2/3}$, and $n$ is the density. Looking at the three fastest quenches, $N_{\mathrm{nG}}/N$, nearly reaches its equilibrium value in less than 1~ms and is essentially stable by 10~ms. Longer quenches show very stable levels of $N_{\mathrm{nG}}$ indicating that pairing is essentially complete by the end of the quench. Higher final values of $N_{\mathrm{nG}}$ for large $t_q$ are consistent with lower final temperatures as less entropy is added during slower quenches.

While the non-Gaussian (paired) fraction builds up quickly, it can take considerably longer for the momentum distribution to reach the expected form for an equilibrium pair condensate. By subtracting the Gaussian fit to the wings of $n_{\mathrm{1D,TOF}}(x)$ from the full distribution, we find the momentum distribution of the non-Gaussian (paired) component $n_{\mathrm{nG}}(x)$. Examples of $n_{\mathrm{nG}}(x)$ for different hold times following a 50~$\mu$s quench are plotted in the insets of Fig.~2(b). These show that even while the area under the curve $N_{\mathrm{nG}}$ is nearly constant, the shape undergoes significant evolution. The main panel of Fig.~2(b) shows the ratio of the rms width $\sigma_{rms}$ of the non-Gaussian component (calculated from the square root of the second moment), relative to the height of the non-Gaussian peak $n_{\mathrm{nG}}(x=0)$. This ratio is sensitive to the shape of the distribution and shows clearly how the momentum distribution becomes sharper with increasing hold time. In contrast to $N_{\mathrm{nG}}$, the shape of $n_{\mathrm{nG}}(x)$ can take $\sim$ 100~ms to fully equilibrate. This reflects the long timescales associated with the dynamics of pairs in low-momentum modes which are most pronounced for the three fastest quenches. Even though the radial monopole mode can be excited with an amplitude of up to a few percent, this does not show up in $N_{\mathrm{nG}}/N$ or the relative width $\sigma_{rms}/n_{\mathrm{nG},0}$, as expected in a scale-invariant system like the unitary Fermi gas. 

For quenches longer than $=\tau_{\mathrm{HO}}$ condensate formation is significantly more adiabatic. When $t_q \geq 2.5$~ms the non-Gaussian component is well-described by an inverted parabola for all $t_h$ and $N_{\mathrm{nG}}/N$ becomes essentially equivalent to a measure of the condensate fraction $N_{0}/N$, obtained from a bi-modal fit to the full distribution $n_{\mathrm{1D,TOF}}(x)$ \cite{Regal04}. This indicates $\tau_{\mathrm{HO}}$ sets the global timescale for condensate formation. Following a quench, condensate growth is expected to seed locally, resulting in the appearance of defects (vortices), according to the Kibble-Zurek mechanism \cite{Ko19,Liu19}. Using shorter expansion times $t_{\mathrm{TOF}} \sim
6$ ms we also observe vortices in clouds following the quench however, their presence does not significantly alter the appearance of the bi-modal momentum distribution at $t_{\mathrm{TOF}} = t_r/4$. The increased widths of the pair momentum distribution in Fig.~2(b) are a result of higher energy excitations, and lead to a behaviour after expansion that appears similar to turbulent regimes in Bose-Einstein condensates \cite{Henn09}. 

Finally, we have measured the thermodynamic properties of the clouds once they have fully equilibrated at unitarity as a function of the quench time. Fig.~3 (blue circles) shows the temperature, determined at unitarity by fitting the 1D-density \emph{in situ} profiles of trapped clouds to the equation of state for the pressure \cite{Ku12}, after a total time $t_q + t_h = 1.4$~s. Here, the temperature is defined relative to the local Fermi temperature $T_F$ in the centre of the clouds where the local density is found using the inverse Abel transform. We observe a clear difference between the final temperatures for slow and fast quenches, with a monotonic decrease in the final relative temperature as the quench time is increased. In the adiabatic limit ($t_q = 400$~ms, where we observe no dynamics after the quench) the relative temperature increases from $T/T_F^{\mathrm{HO}} = 0.06 \pm 0.01$ in the BCS regime to $T/T_F = 0.09 \pm 0.01$ at unitarity. In the BCS limit, $T_F^{\mathrm{HO}} \rightarrow T_F$ at the cloud centre. The data in Fig.~3 indicate a strong change in the adiabaticity for $t_q \sim \tau_{\mathrm{HO}}$ with longer quenches being near-adiabatic, as in Fig. 2(b).

\begin{figure}[ht]
	\centering
		\includegraphics[clip, width = 1 \columnwidth]{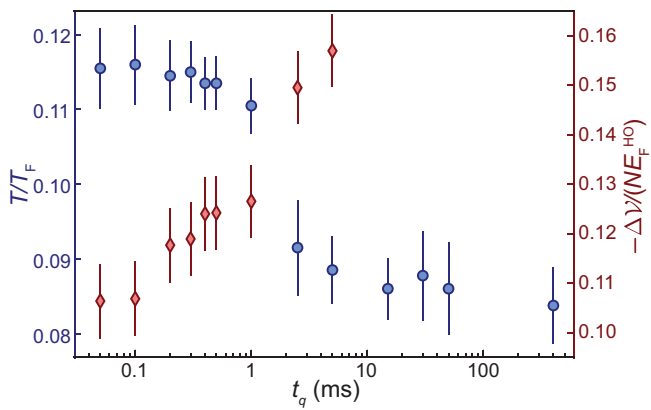}
		\caption{(Color online) Thermodynamic properties of equilibrated clouds for different quench rates. Blue circles show the equilibrium values of the relative temperature $T/T_F$, where $T_F$ is the Fermi temperature in the cloud centre for different quench times after being held for $t_h > 1$ s after the quench. Red diamonds display the right hand side of Eq.~(1), $\Delta {\mathcal V}$, for quenches with $t_q \lesssim \tau_{\mathrm{HO}}$, based on the measured values of the potential energy after equilibration. }
		\label{fig:fig3}
\end{figure}

While sweeping interactions from the BCS regime to unitarity leads to an increase in the relative temperature \cite{Carr04}, the total energy change $\Delta E$ is generally negative and set by the universal contact parameter ${\mathcal I}$. The contact measures the probability of finding two atoms at small separations and governs several thermodynamic and high-momentum properties. It also sets the amplitude of the short-range density-density correlator $n^{(2)}_{\uparrow\downarrow}(r') = {\mathcal I}/(16 \pi^2) [ 1/r'^2 + 2/(r'a) ]$ as $r' \to 0$, where $n^{(2)}_{\uparrow \downarrow}(r') = \int \left < \hat{n}_{\uparrow}(\mathbf{r}-r'/2) \hat{n}_{\downarrow}(\mathbf{r}+r'/2) \right > d^3 \mathbf{r}$ and $\hat{n}_{\sigma}(\mathbf{r})$ is the density operator for atoms in state $\left | \sigma \right >$ \cite{Tan08a}. For an interaction quench $\Delta E$ can be found by integrating the dynamic sweep relation $\dot{E} = - \hbar^2/ (4 \pi m) {\mathcal I(t)} \dot{a^{-1}}(t) + \dot{\langle V(t) \rangle}$, where $\dot{} \equiv d/dt$ represents the time derivative and $\dot{\langle V(t) \rangle}$ accounts for the energy change resulting from a change in the confining potential \cite{Tan08b,Braaten12}. Thus, by examining the energy change for different quenches, we gain insight into the evolution of the contact. To determine the energy before and after the quench, we employ the virial theorem for a harmonic trap \cite{Thomas05,Tan08c}, $E - 2 \langle V \rangle = - {\mathcal I} \hbar^2 / (8 \pi m a)$, where $\langle V \rangle = \int n(\mathbf{r}) V(\mathbf{r}) d^3\mathbf{r} = 3/2 N m \omega_r^2 \langle x^2 \rangle = 3/2 N m \omega_r^2 \langle y^2 \rangle$ is the potential energy. Due to the radial symmetry of our trap, the mean square cloud sizes $\langle x^2 \rangle = \langle y^2 \rangle$ within our experimental uncertainties. At unitarity, this allows straightforward determination of the total energy from the clouds size. However, for weakly interacting clouds the total energy also requires the contact which is \emph{a priori} unknown. Nonetheless, by considering the change in energy resulting from the quench, we can integrate the dynamic sweep theorem by parts and combine this with the virial theorem to show,
\begin{equation}
    \frac{\hbar^2 }{4 \pi m} \left ( \frac{{\mathcal I}_i}{2 a_i} + \int_0^{t_q} \frac{\dot{{\mathcal I}}(t)}{a(t)} dt \right ) = 2 ( \langle V_f \rangle - \langle V_i \rangle) - \Delta V_{\mathrm{HO}}   
\end{equation}
where ${\mathcal I}_i$ and $a_i$ are the contact and scattering length before the quench, $\langle V_i \rangle$ and $\langle V_f \rangle$ are the potential energies before and after the quench. When $t_q$ is short compared to $1/\omega_{\mathrm{HO}} \approx 3$~ms, $n(\mathbf{r})$ will not change significantly during the quench and $\int_0^{t_q} \dot{\langle V(t) \rangle} dt = N m \langle x_i^2 \rangle (\omega_{r_f}^2 - \omega_{r_i}^2) \equiv \Delta V_{\mathrm{HO}}$ where, $\langle x_i^2 \rangle$ is the mean square width of the cloud before the quench and $\omega_{r_i}$ and $\omega_{r_f}$ are the radial trap frequencies before and after the quench, respectively. The right hand side of Eq.~(1) depends upon changes in potential energies and is based entirely on experimentally measured quantities. For convenience, we label this as $\Delta {\mathcal V} \equiv 2 ( \langle V_f \rangle - \langle V_i \rangle) - \Delta V_{\mathrm{HO}}$. In contrast, the left hand side of Eq.~(1) depends on the contact, both before and during the quench. In Fig.~3 (red diamonds) we also plot $\Delta {\mathcal V} / (N E_F^{\mathrm{HO}})$ for $t_q \leq 5$ ms. As expected, the total energy decreases as a result of the quench, but the size of the decrease is smaller for shorter (less adiabatic) quenches. We note that $\Delta V_{\mathrm{HO}}/ (N E_F^{\mathrm{HO}}) \approx -0.03$.

For a quench fast compared to the many-body timescale, the dynamic sweep theorem can be simplified under the assumption that the contact does not change during the quench \cite{Tan08b}. In this limit, the second term on the left hand side of Eq.~(1) vanishes and one could determine the initial contact. However, even for our fastest experimentally accessible quenches we do not reach this limit. To see this, we can estimate the contact at $1/(k_F^{\mathrm{HO}} a_i) = -1.89$ and $T \rightarrow 0$ using the BCS result \cite{Werner09} which yields, ${\mathcal I}_i/(N k_F^{\mathrm{HO}}) \approx 0.14$ and hence ${\mathcal I}_i \hbar^2 / (8 \pi m a_i ) = -0.021 N E_F^{\mathrm{HO}}$. This is significantly smaller than any of the measured values in Fig.~3(b), which indicates that the second term on the left of Eq.~(1) must be $\gtrsim$~4 times larger than the first term, showing that contact undergoes significant evolution during the quench. A similar rapid build up of the contact was seen in a unitary Bose gas \cite{Makotyn14} and also for a Fermi gas near a $p$-wave Feshbach resonance \cite{Luciuk16}. Our shortest quench, $t_q = 50$~$\mu$s, is still longer than the many-body timescale, $\tau_F \approx 26$~$\mu$s in the trap centre and we thus conclude that the short-range (high-momentum) correlations, described by the contact, evolve on even shorter timescales. This is in stark contrast to the long-range (low-momentum) correlations required for the formation of a pair condensate that can take up to 4 orders of magnitude longer to equilibrate.

In summary, we have studied the dynamics of correlations, pair-condensation and the thermodynamic properties of a two-component Fermi gas following an interaction quench to the unitarity limit. At very short timescales, we observe the rapid build up of short-range (high-$k$) correlations, leading to a quick growth of the contact. Over longer timescales, several $\tau_F$, we observe the formation of pairs, signified by the appearance of a central peak in the momentum distribution, with a momentum distribution that becomes narrower over time eventually reaching the shape expected for an equilibrium condensate. The degree of dynamics in low-$k$ modes depends strongly upon the adiabaticity of the quench becoming more pronounced as the quench becomes more non-adiabatic. For quenches long compared to inverse of the lowest trapping frequency $t_q \gg 1/\omega_r$, the quench is effectively adiabatic and isentropically connects different points on the BCS-BEC crossover phase diagram. Our study provides quantitative insight into non-equilibrium dynamics follwoing a quench across the superfluid transition and may be extended to probe universal limits on the rate of energy changes in quantum gases \cite{Qi21}.

We thank Tapio Simula, Meera Parish, Jesper Levinsen and Matthew Davis for helpful discussions and feedback on the manuscript. This research was supported by the ARC Centre of Excellence in Future Low-Energy Electronics Technologies (CE170100039).


\bibliographystyle{apsrev4-2}
\bibliography{quenchbib}

\providecommand{\noopsort}[1]{}\providecommand{\singleletter}[1]{#1}%
\begin{thebibliography}{48}%
\makeatletter
\providecommand \@ifxundefined [1]{%
 \@ifx{#1\undefined}
}%
\providecommand \@ifnum [1]{%
 \ifnum #1\expandafter \@firstoftwo
 \else \expandafter \@secondoftwo
 \fi
}%
\providecommand \@ifx [1]{%
 \ifx #1\expandafter \@firstoftwo
 \else \expandafter \@secondoftwo
 \fi
}%
\providecommand \natexlab [1]{#1}%
\providecommand \enquote  [1]{``#1''}%
\providecommand \bibnamefont  [1]{#1}%
\providecommand \bibfnamefont [1]{#1}%
\providecommand \citenamefont [1]{#1}%
\providecommand \href@noop [0]{\@secondoftwo}%
\providecommand \href [0]{\begingroup \@sanitize@url \@href}%
\providecommand \@href[1]{\@@startlink{#1}\@@href}%
\providecommand \@@href[1]{\endgroup#1\@@endlink}%
\providecommand \@sanitize@url [0]{\catcode `\\12\catcode `\$12\catcode
  `\&12\catcode `\#12\catcode `\^12\catcode `\_12\catcode `\%12\relax}%
\providecommand \@@startlink[1]{}%
\providecommand \@@endlink[0]{}%
\providecommand \url  [0]{\begingroup\@sanitize@url \@url }%
\providecommand \@url [1]{\endgroup\@href {#1}{\urlprefix }}%
\providecommand \urlprefix  [0]{URL }%
\providecommand \Eprint [0]{\href }%
\providecommand \doibase [0]{https://doi.org/}%
\providecommand \selectlanguage [0]{\@gobble}%
\providecommand \bibinfo  [0]{\@secondoftwo}%
\providecommand \bibfield  [0]{\@secondoftwo}%
\providecommand \translation [1]{[#1]}%
\providecommand \BibitemOpen [0]{}%
\providecommand \bibitemStop [0]{}%
\providecommand \bibitemNoStop [0]{.\EOS\space}%
\providecommand \EOS [0]{\spacefactor3000\relax}%
\providecommand \BibitemShut  [1]{\csname bibitem#1\endcsname}%
\let\auto@bib@innerbib\@empty
\bibitem [{\citenamefont {Polkovnikov}\ \emph {et~al.}(2011)\citenamefont
  {Polkovnikov}, \citenamefont {Sengupta}, \citenamefont {Silva},\ and\
  \citenamefont {Vengalattore}}]{Polkovnikov11}%
  \BibitemOpen
  \bibfield  {author} {\bibinfo {author} {\bibfnamefont {A.}~\bibnamefont
  {Polkovnikov}}, \bibinfo {author} {\bibfnamefont {K.}~\bibnamefont
  {Sengupta}}, \bibinfo {author} {\bibfnamefont {A.}~\bibnamefont {Silva}},\
  and\ \bibinfo {author} {\bibfnamefont {M.}~\bibnamefont {Vengalattore}},\
  }\href {https://doi.org/10.1103/RevModPhys.83.863} {\bibfield  {journal}
  {\bibinfo  {journal} {Rev. Mod. Phys.}\ }\textbf {\bibinfo {volume} {83}},\
  \bibinfo {pages} {863} (\bibinfo {year} {2011})}\BibitemShut {NoStop}%
\bibitem [{\citenamefont {Eisert}\ \emph {et~al.}(2015)\citenamefont {Eisert},
  \citenamefont {Friesdorf},\ and\ \citenamefont {Gogolin}}]{Eisert15}%
  \BibitemOpen
  \bibfield  {author} {\bibinfo {author} {\bibfnamefont {J.}~\bibnamefont
  {Eisert}}, \bibinfo {author} {\bibfnamefont {M.}~\bibnamefont {Friesdorf}},\
  and\ \bibinfo {author} {\bibfnamefont {C.}~\bibnamefont {Gogolin}},\
  }\href@noop {} {\bibfield  {journal} {\bibinfo  {journal} {Nature Physics}\
  }\textbf {\bibinfo {volume} {11}},\ \bibinfo {pages} {124} (\bibinfo {year}
  {2015})}\BibitemShut {NoStop}%
\bibitem [{\citenamefont {Fausti}\ \emph {et~al.}(2011)\citenamefont {Fausti},
  \citenamefont {Tobey}, \citenamefont {Dean}, \citenamefont {Kaiser},
  \citenamefont {Dienst}, \citenamefont {Hoffmann}, \citenamefont {Pyon},
  \citenamefont {Takayama}, \citenamefont {Takagi},\ and\ \citenamefont
  {Cavalleri}}]{Fausti11}%
  \BibitemOpen
  \bibfield  {author} {\bibinfo {author} {\bibfnamefont {D.}~\bibnamefont
  {Fausti}}, \bibinfo {author} {\bibfnamefont {R.~I.}\ \bibnamefont {Tobey}},
  \bibinfo {author} {\bibfnamefont {N.}~\bibnamefont {Dean}}, \bibinfo {author}
  {\bibfnamefont {S.}~\bibnamefont {Kaiser}}, \bibinfo {author} {\bibfnamefont
  {A.}~\bibnamefont {Dienst}}, \bibinfo {author} {\bibfnamefont {M.~C.}\
  \bibnamefont {Hoffmann}}, \bibinfo {author} {\bibfnamefont {S.}~\bibnamefont
  {Pyon}}, \bibinfo {author} {\bibfnamefont {T.}~\bibnamefont {Takayama}},
  \bibinfo {author} {\bibfnamefont {H.}~\bibnamefont {Takagi}},\ and\ \bibinfo
  {author} {\bibfnamefont {A.}~\bibnamefont {Cavalleri}},\ }\href
  {https://doi.org/10.1126/science.1197294} {\bibfield  {journal} {\bibinfo
  {journal} {Science}\ }\textbf {\bibinfo {volume} {331}},\ \bibinfo {pages}
  {189} (\bibinfo {year} {2011})}\BibitemShut {NoStop}%
\bibitem [{\citenamefont {Robertson}\ and\ \citenamefont
  {Galitski}(2009)}]{Robertson09}%
  \BibitemOpen
  \bibfield  {author} {\bibinfo {author} {\bibfnamefont {A.}~\bibnamefont
  {Robertson}}\ and\ \bibinfo {author} {\bibfnamefont {V.~M.}\ \bibnamefont
  {Galitski}},\ }\href {https://doi.org/10.1103/PhysRevA.80.063609} {\bibfield
  {journal} {\bibinfo  {journal} {Phys. Rev. A}\ }\textbf {\bibinfo {volume}
  {80}},\ \bibinfo {pages} {063609} (\bibinfo {year} {2009})}\BibitemShut
  {NoStop}%
\bibitem [{\citenamefont {Gring}\ \emph {et~al.}(2012)\citenamefont {Gring},
  \citenamefont {Kuhnert}, \citenamefont {Langen}, \citenamefont {Kitagawa},
  \citenamefont {Rauer}, \citenamefont {Schreitl}, \citenamefont {Mazets},
  \citenamefont {Smith}, \citenamefont {Demler},\ and\ \citenamefont
  {Schmiedmayer}}]{Gring12}%
  \BibitemOpen
  \bibfield  {author} {\bibinfo {author} {\bibfnamefont {M.}~\bibnamefont
  {Gring}}, \bibinfo {author} {\bibfnamefont {M.}~\bibnamefont {Kuhnert}},
  \bibinfo {author} {\bibfnamefont {T.}~\bibnamefont {Langen}}, \bibinfo
  {author} {\bibfnamefont {T.}~\bibnamefont {Kitagawa}}, \bibinfo {author}
  {\bibfnamefont {B.}~\bibnamefont {Rauer}}, \bibinfo {author} {\bibfnamefont
  {M.}~\bibnamefont {Schreitl}}, \bibinfo {author} {\bibfnamefont
  {I.}~\bibnamefont {Mazets}}, \bibinfo {author} {\bibfnamefont {D.~A.}\
  \bibnamefont {Smith}}, \bibinfo {author} {\bibfnamefont {E.}~\bibnamefont
  {Demler}},\ and\ \bibinfo {author} {\bibfnamefont {J.}~\bibnamefont
  {Schmiedmayer}},\ }\href {https://doi.org/10.1126/science.1224953} {\bibfield
   {journal} {\bibinfo  {journal} {Science}\ }\textbf {\bibinfo {volume}
  {337}},\ \bibinfo {pages} {1318} (\bibinfo {year} {2012})}\BibitemShut
  {NoStop}%
\bibitem [{\citenamefont {Foster}\ \emph {et~al.}(2014)\citenamefont {Foster},
  \citenamefont {Gurarie}, \citenamefont {Dzero},\ and\ \citenamefont
  {Yuzbashyan}}]{Foster14}%
  \BibitemOpen
  \bibfield  {author} {\bibinfo {author} {\bibfnamefont {M.~S.}\ \bibnamefont
  {Foster}}, \bibinfo {author} {\bibfnamefont {V.}~\bibnamefont {Gurarie}},
  \bibinfo {author} {\bibfnamefont {M.}~\bibnamefont {Dzero}},\ and\ \bibinfo
  {author} {\bibfnamefont {E.~A.}\ \bibnamefont {Yuzbashyan}},\ }\href
  {https://doi.org/10.1103/PhysRevLett.113.076403} {\bibfield  {journal}
  {\bibinfo  {journal} {Phys. Rev. Lett.}\ }\textbf {\bibinfo {volume} {113}},\
  \bibinfo {pages} {076403} (\bibinfo {year} {2014})}\BibitemShut {NoStop}%
\bibitem [{\citenamefont {Miesner}\ \emph {et~al.}(1998)\citenamefont
  {Miesner}, \citenamefont {Stamper-Kurn}, \citenamefont {Andrews},
  \citenamefont {Durfee}, \citenamefont {Inouye},\ and\ \citenamefont
  {Ketterle}}]{Miesner98}%
  \BibitemOpen
  \bibfield  {author} {\bibinfo {author} {\bibfnamefont {H.-J.}\ \bibnamefont
  {Miesner}}, \bibinfo {author} {\bibfnamefont {D.~M.}\ \bibnamefont
  {Stamper-Kurn}}, \bibinfo {author} {\bibfnamefont {M.~R.}\ \bibnamefont
  {Andrews}}, \bibinfo {author} {\bibfnamefont {D.~S.}\ \bibnamefont {Durfee}},
  \bibinfo {author} {\bibfnamefont {S.}~\bibnamefont {Inouye}},\ and\ \bibinfo
  {author} {\bibfnamefont {W.}~\bibnamefont {Ketterle}},\ }\href
  {https://doi.org/10.1126/science.279.5353.1005} {\bibfield  {journal}
  {\bibinfo  {journal} {Science}\ }\textbf {\bibinfo {volume} {279}},\ \bibinfo
  {pages} {1005} (\bibinfo {year} {1998})}\BibitemShut {NoStop}%
\bibitem [{\citenamefont {K\"ohl}\ \emph {et~al.}(2002)\citenamefont {K\"ohl},
  \citenamefont {Davis}, \citenamefont {Gardiner}, \citenamefont {H\"ansch},\
  and\ \citenamefont {Esslinger}}]{Kohl04}%
  \BibitemOpen
  \bibfield  {author} {\bibinfo {author} {\bibfnamefont {M.}~\bibnamefont
  {K\"ohl}}, \bibinfo {author} {\bibfnamefont {M.~J.}\ \bibnamefont {Davis}},
  \bibinfo {author} {\bibfnamefont {C.~W.}\ \bibnamefont {Gardiner}}, \bibinfo
  {author} {\bibfnamefont {T.~W.}\ \bibnamefont {H\"ansch}},\ and\ \bibinfo
  {author} {\bibfnamefont {T.}~\bibnamefont {Esslinger}},\ }\href
  {https://doi.org/10.1103/PhysRevLett.88.080402} {\bibfield  {journal}
  {\bibinfo  {journal} {Phys. Rev. Lett.}\ }\textbf {\bibinfo {volume} {88}},\
  \bibinfo {pages} {080402} (\bibinfo {year} {2002})}\BibitemShut {NoStop}%
\bibitem [{\citenamefont {Smith}\ \emph {et~al.}(2012)\citenamefont {Smith},
  \citenamefont {Beattie}, \citenamefont {Moulder}, \citenamefont {Campbell},\
  and\ \citenamefont {Hadzibabic}}]{Smith12}%
  \BibitemOpen
  \bibfield  {author} {\bibinfo {author} {\bibfnamefont {R.~P.}\ \bibnamefont
  {Smith}}, \bibinfo {author} {\bibfnamefont {S.}~\bibnamefont {Beattie}},
  \bibinfo {author} {\bibfnamefont {S.}~\bibnamefont {Moulder}}, \bibinfo
  {author} {\bibfnamefont {R.~L.~D.}\ \bibnamefont {Campbell}},\ and\ \bibinfo
  {author} {\bibfnamefont {Z.}~\bibnamefont {Hadzibabic}},\ }\href
  {https://doi.org/10.1103/PhysRevLett.109.105301} {\bibfield  {journal}
  {\bibinfo  {journal} {Phys. Rev. Lett.}\ }\textbf {\bibinfo {volume} {109}},\
  \bibinfo {pages} {105301} (\bibinfo {year} {2012})}\BibitemShut {NoStop}%
\bibitem [{\citenamefont {Davis}\ \emph {et~al.}(2017)\citenamefont {Davis},
  \citenamefont {Wright}, \citenamefont {Gasenzer}, \citenamefont {Gardiner},\
  and\ \citenamefont {Proukakis}}]{Davis17}%
  \BibitemOpen
  \bibfield  {author} {\bibinfo {author} {\bibfnamefont {M.~J.}\ \bibnamefont
  {Davis}}, \bibinfo {author} {\bibfnamefont {T.~M.}\ \bibnamefont {Wright}},
  \bibinfo {author} {\bibfnamefont {T.}~\bibnamefont {Gasenzer}}, \bibinfo
  {author} {\bibfnamefont {S.~A.}\ \bibnamefont {Gardiner}},\ and\ \bibinfo
  {author} {\bibfnamefont {N.~P.}\ \bibnamefont {Proukakis}},\ }\bibinfo
  {title} {{Formation of Bose-Einstein Condensates}},\ in\ \href
  {https://doi.org/10.1017/9781316084366.009} {\emph {\bibinfo {booktitle}
  {Universal Themes of Bose-Einstein Condensation}}},\ \bibinfo {editor}
  {edited by\ \bibinfo {editor} {\bibfnamefont {N.~P.}\ \bibnamefont
  {Proukakis}}, \bibinfo {editor} {\bibfnamefont {D.~W.}\ \bibnamefont
  {Snoke}},\ and\ \bibinfo {editor} {\bibfnamefont {P.~B.}\ \bibnamefont
  {Littlewood}}}\ (\bibinfo  {publisher} {Cambridge University Press},\
  \bibinfo {year} {2017})\ pp.\ \bibinfo {pages} {117--150}\BibitemShut
  {NoStop}%
\bibitem [{\citenamefont {Papp}\ \emph {et~al.}(2008)\citenamefont {Papp},
  \citenamefont {Pino}, \citenamefont {Wild}, \citenamefont {Ronen},
  \citenamefont {Wieman}, \citenamefont {Jin},\ and\ \citenamefont
  {Cornell}}]{Papp08}%
  \BibitemOpen
  \bibfield  {author} {\bibinfo {author} {\bibfnamefont {S.~B.}\ \bibnamefont
  {Papp}}, \bibinfo {author} {\bibfnamefont {J.~M.}\ \bibnamefont {Pino}},
  \bibinfo {author} {\bibfnamefont {R.~J.}\ \bibnamefont {Wild}}, \bibinfo
  {author} {\bibfnamefont {S.}~\bibnamefont {Ronen}}, \bibinfo {author}
  {\bibfnamefont {C.~E.}\ \bibnamefont {Wieman}}, \bibinfo {author}
  {\bibfnamefont {D.~S.}\ \bibnamefont {Jin}},\ and\ \bibinfo {author}
  {\bibfnamefont {E.~A.}\ \bibnamefont {Cornell}},\ }\href
  {https://doi.org/10.1103/PhysRevLett.101.135301} {\bibfield  {journal}
  {\bibinfo  {journal} {Phys. Rev. Lett.}\ }\textbf {\bibinfo {volume} {101}},\
  \bibinfo {pages} {135301} (\bibinfo {year} {2008})}\BibitemShut {NoStop}%
\bibitem [{\citenamefont {Lopes}\ \emph {et~al.}(2017)\citenamefont {Lopes},
  \citenamefont {Eigen}, \citenamefont {Barker}, \citenamefont {Viebahn},
  \citenamefont {Robert-de Saint-Vincent}, \citenamefont {Navon}, \citenamefont
  {Hadzibabic},\ and\ \citenamefont {Smith}}]{Lopes17a}%
  \BibitemOpen
  \bibfield  {author} {\bibinfo {author} {\bibfnamefont {R.}~\bibnamefont
  {Lopes}}, \bibinfo {author} {\bibfnamefont {C.}~\bibnamefont {Eigen}},
  \bibinfo {author} {\bibfnamefont {A.}~\bibnamefont {Barker}}, \bibinfo
  {author} {\bibfnamefont {K.~G.~H.}\ \bibnamefont {Viebahn}}, \bibinfo
  {author} {\bibfnamefont {M.}~\bibnamefont {Robert-de Saint-Vincent}},
  \bibinfo {author} {\bibfnamefont {N.}~\bibnamefont {Navon}}, \bibinfo
  {author} {\bibfnamefont {Z.}~\bibnamefont {Hadzibabic}},\ and\ \bibinfo
  {author} {\bibfnamefont {R.~P.}\ \bibnamefont {Smith}},\ }\href
  {https://doi.org/10.1103/PhysRevLett.118.210401} {\bibfield  {journal}
  {\bibinfo  {journal} {Phys. Rev. Lett.}\ }\textbf {\bibinfo {volume} {118}},\
  \bibinfo {pages} {210401} (\bibinfo {year} {2017})}\BibitemShut {NoStop}%
\bibitem [{\citenamefont {Weiler}\ \emph {et~al.}(2008)\citenamefont {Weiler},
  \citenamefont {Neely}, \citenamefont {Scherer}, \citenamefont {Bradley},
  \citenamefont {Davis},\ and\ \citenamefont {Anderson}}]{Weiler08}%
  \BibitemOpen
  \bibfield  {author} {\bibinfo {author} {\bibfnamefont {C.~N.}\ \bibnamefont
  {Weiler}}, \bibinfo {author} {\bibfnamefont {T.~W.}\ \bibnamefont {Neely}},
  \bibinfo {author} {\bibfnamefont {D.~R.}\ \bibnamefont {Scherer}}, \bibinfo
  {author} {\bibfnamefont {A.~S.}\ \bibnamefont {Bradley}}, \bibinfo {author}
  {\bibfnamefont {M.~J.}\ \bibnamefont {Davis}},\ and\ \bibinfo {author}
  {\bibfnamefont {B.~P.}\ \bibnamefont {Anderson}},\ }\href@noop {} {\bibfield
  {journal} {\bibinfo  {journal} {Nature}\ }\textbf {\bibinfo {volume} {455}},\
  \bibinfo {pages} {948} (\bibinfo {year} {2008})}\BibitemShut {NoStop}%
\bibitem [{\citenamefont {Wild}\ \emph {et~al.}(2012)\citenamefont {Wild},
  \citenamefont {Makotyn}, \citenamefont {Pino}, \citenamefont {Cornell},\ and\
  \citenamefont {Jin}}]{Wild12}%
  \BibitemOpen
  \bibfield  {author} {\bibinfo {author} {\bibfnamefont {R.~J.}\ \bibnamefont
  {Wild}}, \bibinfo {author} {\bibfnamefont {P.}~\bibnamefont {Makotyn}},
  \bibinfo {author} {\bibfnamefont {J.~M.}\ \bibnamefont {Pino}}, \bibinfo
  {author} {\bibfnamefont {E.~A.}\ \bibnamefont {Cornell}},\ and\ \bibinfo
  {author} {\bibfnamefont {D.~S.}\ \bibnamefont {Jin}},\ }\href
  {https://doi.org/10.1103/PhysRevLett.108.145305} {\bibfield  {journal}
  {\bibinfo  {journal} {Phys. Rev. Lett.}\ }\textbf {\bibinfo {volume} {108}},\
  \bibinfo {pages} {145305} (\bibinfo {year} {2012})}\BibitemShut {NoStop}%
\bibitem [{\citenamefont {Makotyn}\ \emph {et~al.}(2014)\citenamefont
  {Makotyn}, \citenamefont {Klauss}, \citenamefont {Goldberger}, \citenamefont
  {Cornell},\ and\ \citenamefont {Jin}}]{Makotyn14}%
  \BibitemOpen
  \bibfield  {author} {\bibinfo {author} {\bibfnamefont {P.}~\bibnamefont
  {Makotyn}}, \bibinfo {author} {\bibfnamefont {C.~E.}\ \bibnamefont {Klauss}},
  \bibinfo {author} {\bibfnamefont {D.~L.}\ \bibnamefont {Goldberger}},
  \bibinfo {author} {\bibfnamefont {E.}~\bibnamefont {Cornell}},\ and\ \bibinfo
  {author} {\bibfnamefont {D.~S.}\ \bibnamefont {Jin}},\ }\href@noop {}
  {\bibfield  {journal} {\bibinfo  {journal} {Nature Physics}\ }\textbf
  {\bibinfo {volume} {10}},\ \bibinfo {pages} {116} (\bibinfo {year}
  {2014})}\BibitemShut {NoStop}%
\bibitem [{\citenamefont {Fletcher}\ \emph {et~al.}(2017)\citenamefont
  {Fletcher}, \citenamefont {Lopes}, \citenamefont {Man}, \citenamefont
  {Navon}, \citenamefont {Smith}, \citenamefont {Zwierlein},\ and\
  \citenamefont {Hadzibabic}}]{Fletcher17}%
  \BibitemOpen
  \bibfield  {author} {\bibinfo {author} {\bibfnamefont {R.~J.}\ \bibnamefont
  {Fletcher}}, \bibinfo {author} {\bibfnamefont {R.}~\bibnamefont {Lopes}},
  \bibinfo {author} {\bibfnamefont {J.}~\bibnamefont {Man}}, \bibinfo {author}
  {\bibfnamefont {N.}~\bibnamefont {Navon}}, \bibinfo {author} {\bibfnamefont
  {R.~P.}\ \bibnamefont {Smith}}, \bibinfo {author} {\bibfnamefont {M.~W.}\
  \bibnamefont {Zwierlein}},\ and\ \bibinfo {author} {\bibfnamefont
  {Z.}~\bibnamefont {Hadzibabic}},\ }\href@noop {} {\bibfield  {journal}
  {\bibinfo  {journal} {Science}\ }\textbf {\bibinfo {volume} {355}},\ \bibinfo
  {pages} {377} (\bibinfo {year} {2017})}\BibitemShut {NoStop}%
\bibitem [{\citenamefont {Erne}\ \emph {et~al.}(2018)\citenamefont {Erne},
  \citenamefont {B{\"u}cker}, \citenamefont {Gasenzer}, \citenamefont
  {Berges},\ and\ \citenamefont {Schmiedmayer}}]{Erne18}%
  \BibitemOpen
  \bibfield  {author} {\bibinfo {author} {\bibfnamefont {S.}~\bibnamefont
  {Erne}}, \bibinfo {author} {\bibfnamefont {R.}~\bibnamefont {B{\"u}cker}},
  \bibinfo {author} {\bibfnamefont {T.}~\bibnamefont {Gasenzer}}, \bibinfo
  {author} {\bibfnamefont {J.}~\bibnamefont {Berges}},\ and\ \bibinfo {author}
  {\bibfnamefont {J.}~\bibnamefont {Schmiedmayer}},\ }\href@noop {} {\bibfield
  {journal} {\bibinfo  {journal} {Nature}\ }\textbf {\bibinfo {volume} {563}},\
  \bibinfo {pages} {225} (\bibinfo {year} {2018})}\BibitemShut {NoStop}%
\bibitem [{\citenamefont {Eigen}\ \emph {et~al.}(2018)\citenamefont {Eigen},
  \citenamefont {Glidden}, \citenamefont {Lopes}, \citenamefont {Cornell},
  \citenamefont {Smith},\ and\ \citenamefont {Hadzibabic}}]{Eigen18}%
  \BibitemOpen
  \bibfield  {author} {\bibinfo {author} {\bibfnamefont {C.}~\bibnamefont
  {Eigen}}, \bibinfo {author} {\bibfnamefont {J.~A.}\ \bibnamefont {Glidden}},
  \bibinfo {author} {\bibfnamefont {R.}~\bibnamefont {Lopes}}, \bibinfo
  {author} {\bibfnamefont {E.~A.}\ \bibnamefont {Cornell}}, \bibinfo {author}
  {\bibfnamefont {R.~P.}\ \bibnamefont {Smith}},\ and\ \bibinfo {author}
  {\bibfnamefont {Z.}~\bibnamefont {Hadzibabic}},\ }\href@noop {} {\bibfield
  {journal} {\bibinfo  {journal} {Nature}\ }\textbf {\bibinfo {volume} {563}},\
  \bibinfo {pages} {221} (\bibinfo {year} {2018})}\BibitemShut {NoStop}%
\bibitem [{\citenamefont {Pr{\"u}fer}\ \emph {et~al.}(2018)\citenamefont
  {Pr{\"u}fer}, \citenamefont {Kunkel}, \citenamefont {Strobel}, \citenamefont
  {Lannig}, \citenamefont {Linnemann}, \citenamefont {Schmied}, \citenamefont
  {Berges}, \citenamefont {Gasenzer},\ and\ \citenamefont
  {Oberthaler}}]{Prufer18}%
  \BibitemOpen
  \bibfield  {author} {\bibinfo {author} {\bibfnamefont {M.}~\bibnamefont
  {Pr{\"u}fer}}, \bibinfo {author} {\bibfnamefont {P.}~\bibnamefont {Kunkel}},
  \bibinfo {author} {\bibfnamefont {H.}~\bibnamefont {Strobel}}, \bibinfo
  {author} {\bibfnamefont {S.}~\bibnamefont {Lannig}}, \bibinfo {author}
  {\bibfnamefont {D.}~\bibnamefont {Linnemann}}, \bibinfo {author}
  {\bibfnamefont {C.-M.}\ \bibnamefont {Schmied}}, \bibinfo {author}
  {\bibfnamefont {J.}~\bibnamefont {Berges}}, \bibinfo {author} {\bibfnamefont
  {T.}~\bibnamefont {Gasenzer}},\ and\ \bibinfo {author} {\bibfnamefont
  {M.~K.}\ \bibnamefont {Oberthaler}},\ }\href@noop {} {\bibfield  {journal}
  {\bibinfo  {journal} {Nature}\ }\textbf {\bibinfo {volume} {563}},\ \bibinfo
  {pages} {217} (\bibinfo {year} {2018})}\BibitemShut {NoStop}%
\bibitem [{\citenamefont {Rem}\ \emph {et~al.}(2013)\citenamefont {Rem},
  \citenamefont {Grier}, \citenamefont {Ferrier-Barbut}, \citenamefont
  {Eismann}, \citenamefont {Langen}, \citenamefont {Navon}, \citenamefont
  {Khaykovich}, \citenamefont {Werner}, \citenamefont {Petrov}, \citenamefont
  {Chevy},\ and\ \citenamefont {Salomon}}]{Rem13}%
  \BibitemOpen
  \bibfield  {author} {\bibinfo {author} {\bibfnamefont {B.~S.}\ \bibnamefont
  {Rem}}, \bibinfo {author} {\bibfnamefont {A.~T.}\ \bibnamefont {Grier}},
  \bibinfo {author} {\bibfnamefont {I.}~\bibnamefont {Ferrier-Barbut}},
  \bibinfo {author} {\bibfnamefont {U.}~\bibnamefont {Eismann}}, \bibinfo
  {author} {\bibfnamefont {T.}~\bibnamefont {Langen}}, \bibinfo {author}
  {\bibfnamefont {N.}~\bibnamefont {Navon}}, \bibinfo {author} {\bibfnamefont
  {L.}~\bibnamefont {Khaykovich}}, \bibinfo {author} {\bibfnamefont
  {F.}~\bibnamefont {Werner}}, \bibinfo {author} {\bibfnamefont {D.~S.}\
  \bibnamefont {Petrov}}, \bibinfo {author} {\bibfnamefont {F.}~\bibnamefont
  {Chevy}},\ and\ \bibinfo {author} {\bibfnamefont {C.}~\bibnamefont
  {Salomon}},\ }\href {https://doi.org/10.1103/PhysRevLett.110.163202}
  {\bibfield  {journal} {\bibinfo  {journal} {Phys. Rev. Lett.}\ }\textbf
  {\bibinfo {volume} {110}},\ \bibinfo {pages} {163202} (\bibinfo {year}
  {2013})}\BibitemShut {NoStop}%
\bibitem [{\citenamefont {Fletcher}\ \emph {et~al.}(2013)\citenamefont
  {Fletcher}, \citenamefont {Gaunt}, \citenamefont {Navon}, \citenamefont
  {Smith},\ and\ \citenamefont {Hadzibabic}}]{Fletcher13}%
  \BibitemOpen
  \bibfield  {author} {\bibinfo {author} {\bibfnamefont {R.~J.}\ \bibnamefont
  {Fletcher}}, \bibinfo {author} {\bibfnamefont {A.~L.}\ \bibnamefont {Gaunt}},
  \bibinfo {author} {\bibfnamefont {N.}~\bibnamefont {Navon}}, \bibinfo
  {author} {\bibfnamefont {R.~P.}\ \bibnamefont {Smith}},\ and\ \bibinfo
  {author} {\bibfnamefont {Z.}~\bibnamefont {Hadzibabic}},\ }\href
  {https://doi.org/10.1103/PhysRevLett.111.125303} {\bibfield  {journal}
  {\bibinfo  {journal} {Phys. Rev. Lett.}\ }\textbf {\bibinfo {volume} {111}},\
  \bibinfo {pages} {125303} (\bibinfo {year} {2013})}\BibitemShut {NoStop}%
\bibitem [{\citenamefont {Petrov}\ \emph {et~al.}(2004)\citenamefont {Petrov},
  \citenamefont {Salomon},\ and\ \citenamefont {Shlyapnikov}}]{Petrov04}%
  \BibitemOpen
  \bibfield  {author} {\bibinfo {author} {\bibfnamefont {D.~S.}\ \bibnamefont
  {Petrov}}, \bibinfo {author} {\bibfnamefont {C.}~\bibnamefont {Salomon}},\
  and\ \bibinfo {author} {\bibfnamefont {G.~V.}\ \bibnamefont {Shlyapnikov}},\
  }\href {https://doi.org/10.1103/PhysRevLett.93.090404} {\bibfield  {journal}
  {\bibinfo  {journal} {Phys. Rev. Lett.}\ }\textbf {\bibinfo {volume} {93}},\
  \bibinfo {pages} {090404} (\bibinfo {year} {2004})}\BibitemShut {NoStop}%
\bibitem [{\citenamefont {Zwierlein}\ \emph {et~al.}(2005)\citenamefont
  {Zwierlein}, \citenamefont {Schunck}, \citenamefont {Stan}, \citenamefont
  {Raupach},\ and\ \citenamefont {Ketterle}}]{Zwierlein05}%
  \BibitemOpen
  \bibfield  {author} {\bibinfo {author} {\bibfnamefont {M.~W.}\ \bibnamefont
  {Zwierlein}}, \bibinfo {author} {\bibfnamefont {C.~H.}\ \bibnamefont
  {Schunck}}, \bibinfo {author} {\bibfnamefont {C.~A.}\ \bibnamefont {Stan}},
  \bibinfo {author} {\bibfnamefont {S.~M.~F.}\ \bibnamefont {Raupach}},\ and\
  \bibinfo {author} {\bibfnamefont {W.}~\bibnamefont {Ketterle}},\ }\href
  {https://doi.org/10.1103/PhysRevLett.94.180401} {\bibfield  {journal}
  {\bibinfo  {journal} {Phys. Rev. Lett.}\ }\textbf {\bibinfo {volume} {94}},\
  \bibinfo {pages} {180401} (\bibinfo {year} {2005})}\BibitemShut {NoStop}%
\bibitem [{\citenamefont {Ko}\ \emph {et~al.}(2019)\citenamefont {Ko},
  \citenamefont {Park},\ and\ \citenamefont {Shin}}]{Ko19}%
  \BibitemOpen
  \bibfield  {author} {\bibinfo {author} {\bibfnamefont {B.}~\bibnamefont
  {Ko}}, \bibinfo {author} {\bibfnamefont {J.~W.}\ \bibnamefont {Park}},\ and\
  \bibinfo {author} {\bibfnamefont {Y.-i.}\ \bibnamefont {Shin}},\ }\href@noop
  {} {\bibfield  {journal} {\bibinfo  {journal} {Nature Physics}\ }\textbf
  {\bibinfo {volume} {15}},\ \bibinfo {pages} {1227} (\bibinfo {year}
  {2019})}\BibitemShut {NoStop}%
\bibitem [{\citenamefont {Liu}\ \emph {et~al.}(2019)\citenamefont {Liu},
  \citenamefont {Yao}, \citenamefont {Deng}, \citenamefont {Wang},
  \citenamefont {Wang}, \citenamefont {Li}, \citenamefont {Chen},\ and\
  \citenamefont {Pan}}]{Liu19}%
  \BibitemOpen
  \bibfield  {author} {\bibinfo {author} {\bibfnamefont {X.-P.}\ \bibnamefont
  {Liu}}, \bibinfo {author} {\bibfnamefont {X.-C.}\ \bibnamefont {Yao}},
  \bibinfo {author} {\bibfnamefont {Y.}~\bibnamefont {Deng}}, \bibinfo {author}
  {\bibfnamefont {X.-Q.}\ \bibnamefont {Wang}}, \bibinfo {author}
  {\bibfnamefont {Y.-X.}\ \bibnamefont {Wang}}, \bibinfo {author}
  {\bibfnamefont {X.-P.}\ \bibnamefont {Li}}, \bibinfo {author} {\bibfnamefont
  {Y.-A.}\ \bibnamefont {Chen}},\ and\ \bibinfo {author} {\bibfnamefont
  {J.-W.}\ \bibnamefont {Pan}},\ }\href@noop {} {\bibfield  {journal} {\bibinfo
   {journal} {arXiv preprint arXiv:1902.07558}\ } (\bibinfo {year}
  {2019})}\BibitemShut {NoStop}%
\bibitem [{\citenamefont {Behrle}\ \emph {et~al.}(2018)\citenamefont {Behrle},
  \citenamefont {Harrison}, \citenamefont {Kombe}, \citenamefont {Gao},
  \citenamefont {Link}, \citenamefont {Bernier}, \citenamefont {Kollath},\ and\
  \citenamefont {K{\"o}hl}}]{Behrle18}%
  \BibitemOpen
  \bibfield  {author} {\bibinfo {author} {\bibfnamefont {A.}~\bibnamefont
  {Behrle}}, \bibinfo {author} {\bibfnamefont {T.}~\bibnamefont {Harrison}},
  \bibinfo {author} {\bibfnamefont {J.}~\bibnamefont {Kombe}}, \bibinfo
  {author} {\bibfnamefont {K.}~\bibnamefont {Gao}}, \bibinfo {author}
  {\bibfnamefont {M.}~\bibnamefont {Link}}, \bibinfo {author} {\bibfnamefont
  {J.-S.}\ \bibnamefont {Bernier}}, \bibinfo {author} {\bibfnamefont
  {C.}~\bibnamefont {Kollath}},\ and\ \bibinfo {author} {\bibfnamefont
  {M.}~\bibnamefont {K{\"o}hl}},\ }\href@noop {} {\bibfield  {journal}
  {\bibinfo  {journal} {Nature Physics}\ }\textbf {\bibinfo {volume} {14}},\
  \bibinfo {pages} {781} (\bibinfo {year} {2018})}\BibitemShut {NoStop}%
\bibitem [{\citenamefont {Harrison}\ \emph {et~al.}(2020)\citenamefont
  {Harrison}, \citenamefont {Link}, \citenamefont {Behrle}, \citenamefont
  {Gao}, \citenamefont {Kell}, \citenamefont {Kombe}, \citenamefont {Bernier},
  \citenamefont {Kollath},\ and\ \citenamefont {K\"{o}hl}}]{Harrison20}%
  \BibitemOpen
  \bibfield  {author} {\bibinfo {author} {\bibfnamefont {T.}~\bibnamefont
  {Harrison}}, \bibinfo {author} {\bibfnamefont {M.}~\bibnamefont {Link}},
  \bibinfo {author} {\bibfnamefont {A.}~\bibnamefont {Behrle}}, \bibinfo
  {author} {\bibfnamefont {K.}~\bibnamefont {Gao}}, \bibinfo {author}
  {\bibfnamefont {A.}~\bibnamefont {Kell}}, \bibinfo {author} {\bibfnamefont
  {J.}~\bibnamefont {Kombe}}, \bibinfo {author} {\bibfnamefont {J.-S.}\
  \bibnamefont {Bernier}}, \bibinfo {author} {\bibfnamefont {C.}~\bibnamefont
  {Kollath}},\ and\ \bibinfo {author} {\bibfnamefont {M.}~\bibnamefont
  {K\"{o}hl}},\ }\href@noop {} {\bibinfo {title} {Decay and revival of a
  transient trapped fermi condensate}} (\bibinfo {year} {2020}),\ \Eprint
  {https://arxiv.org/abs/2007.11466} {arXiv:2007.11466 [cond-mat.quant-gas]}
  \BibitemShut {NoStop}%
\bibitem [{\citenamefont {Tan}(2008{\natexlab{a}})}]{Tan08b}%
  \BibitemOpen
  \bibfield  {author} {\bibinfo {author} {\bibfnamefont {S.}~\bibnamefont
  {Tan}},\ }\href@noop {} {\bibfield  {journal} {\bibinfo  {journal} {Annals of
  Physics}\ }\textbf {\bibinfo {volume} {323}},\ \bibinfo {pages} {2971}
  (\bibinfo {year} {2008}{\natexlab{a}})}\BibitemShut {NoStop}%
\bibitem [{\citenamefont {Tan}(2008{\natexlab{b}})}]{Tan08c}%
  \BibitemOpen
  \bibfield  {author} {\bibinfo {author} {\bibfnamefont {S.}~\bibnamefont
  {Tan}},\ }\href@noop {} {\bibfield  {journal} {\bibinfo  {journal} {Annals of
  Physics}\ }\textbf {\bibinfo {volume} {323}},\ \bibinfo {pages} {2987}
  (\bibinfo {year} {2008}{\natexlab{b}})}\BibitemShut {NoStop}%
\bibitem [{\citenamefont {Z\"urn}\ \emph {et~al.}(2013)\citenamefont {Z\"urn},
  \citenamefont {Lompe}, \citenamefont {Wenz}, \citenamefont {Jochim},
  \citenamefont {Julienne},\ and\ \citenamefont {Hutson}}]{Zurn13}%
  \BibitemOpen
  \bibfield  {author} {\bibinfo {author} {\bibfnamefont {G.}~\bibnamefont
  {Z\"urn}}, \bibinfo {author} {\bibfnamefont {T.}~\bibnamefont {Lompe}},
  \bibinfo {author} {\bibfnamefont {A.~N.}\ \bibnamefont {Wenz}}, \bibinfo
  {author} {\bibfnamefont {S.}~\bibnamefont {Jochim}}, \bibinfo {author}
  {\bibfnamefont {P.~S.}\ \bibnamefont {Julienne}},\ and\ \bibinfo {author}
  {\bibfnamefont {J.~M.}\ \bibnamefont {Hutson}},\ }\href
  {https://doi.org/10.1103/PhysRevLett.110.135301} {\bibfield  {journal}
  {\bibinfo  {journal} {Phys. Rev. Lett.}\ }\textbf {\bibinfo {volume} {110}},\
  \bibinfo {pages} {135301} (\bibinfo {year} {2013})}\BibitemShut {NoStop}%
\bibitem [{\citenamefont {Smith}\ \emph {et~al.}(2005)\citenamefont {Smith},
  \citenamefont {Heathcote}, \citenamefont {Hechenblaikner}, \citenamefont
  {Nugent},\ and\ \citenamefont {Foot}}]{Smith05}%
  \BibitemOpen
  \bibfield  {author} {\bibinfo {author} {\bibfnamefont {N.~L.}\ \bibnamefont
  {Smith}}, \bibinfo {author} {\bibfnamefont {W.~H.}\ \bibnamefont
  {Heathcote}}, \bibinfo {author} {\bibfnamefont {G.}~\bibnamefont
  {Hechenblaikner}}, \bibinfo {author} {\bibfnamefont {E.}~\bibnamefont
  {Nugent}},\ and\ \bibinfo {author} {\bibfnamefont {C.~J.}\ \bibnamefont
  {Foot}},\ }\href {https://doi.org/10.1088/0953-4075/38/3/007} {\bibfield
  {journal} {\bibinfo  {journal} {Journal of Physics B: Atomic, Molecular and
  Optical Physics}\ }\textbf {\bibinfo {volume} {38}},\ \bibinfo {pages} {223}
  (\bibinfo {year} {2005})}\BibitemShut {NoStop}%
\bibitem [{\citenamefont {Rath}\ \emph {et~al.}(2010)\citenamefont {Rath},
  \citenamefont {Yefsah}, \citenamefont {G\"unter}, \citenamefont {Cheneau},
  \citenamefont {Desbuquois}, \citenamefont {Holzmann}, \citenamefont
  {Krauth},\ and\ \citenamefont {Dalibard}}]{Rath10}%
  \BibitemOpen
  \bibfield  {author} {\bibinfo {author} {\bibfnamefont {S.~P.}\ \bibnamefont
  {Rath}}, \bibinfo {author} {\bibfnamefont {T.}~\bibnamefont {Yefsah}},
  \bibinfo {author} {\bibfnamefont {K.~J.}\ \bibnamefont {G\"unter}}, \bibinfo
  {author} {\bibfnamefont {M.}~\bibnamefont {Cheneau}}, \bibinfo {author}
  {\bibfnamefont {R.}~\bibnamefont {Desbuquois}}, \bibinfo {author}
  {\bibfnamefont {M.}~\bibnamefont {Holzmann}}, \bibinfo {author}
  {\bibfnamefont {W.}~\bibnamefont {Krauth}},\ and\ \bibinfo {author}
  {\bibfnamefont {J.}~\bibnamefont {Dalibard}},\ }\href
  {https://doi.org/10.1103/PhysRevA.82.013609} {\bibfield  {journal} {\bibinfo
  {journal} {Phys. Rev. A}\ }\textbf {\bibinfo {volume} {82}},\ \bibinfo
  {pages} {013609} (\bibinfo {year} {2010})}\BibitemShut {NoStop}%
\bibitem [{\citenamefont {Dyke}\ \emph {et~al.}(2016)\citenamefont {Dyke},
  \citenamefont {Fenech}, \citenamefont {Peppler}, \citenamefont {Lingham},
  \citenamefont {Hoinka}, \citenamefont {Zhang}, \citenamefont {Peng},
  \citenamefont {Mulkerin}, \citenamefont {Hu}, \citenamefont {Liu},\ and\
  \citenamefont {Vale}}]{Dyke16}%
  \BibitemOpen
  \bibfield  {author} {\bibinfo {author} {\bibfnamefont {P.}~\bibnamefont
  {Dyke}}, \bibinfo {author} {\bibfnamefont {K.}~\bibnamefont {Fenech}},
  \bibinfo {author} {\bibfnamefont {T.}~\bibnamefont {Peppler}}, \bibinfo
  {author} {\bibfnamefont {M.~G.}\ \bibnamefont {Lingham}}, \bibinfo {author}
  {\bibfnamefont {S.}~\bibnamefont {Hoinka}}, \bibinfo {author} {\bibfnamefont
  {W.}~\bibnamefont {Zhang}}, \bibinfo {author} {\bibfnamefont {S.-G.}\
  \bibnamefont {Peng}}, \bibinfo {author} {\bibfnamefont {B.}~\bibnamefont
  {Mulkerin}}, \bibinfo {author} {\bibfnamefont {H.}~\bibnamefont {Hu}},
  \bibinfo {author} {\bibfnamefont {X.-J.}\ \bibnamefont {Liu}},\ and\ \bibinfo
  {author} {\bibfnamefont {C.~J.}\ \bibnamefont {Vale}},\ }\href
  {https://doi.org/10.1103/PhysRevA.93.011603} {\bibfield  {journal} {\bibinfo
  {journal} {Phys. Rev. A}\ }\textbf {\bibinfo {volume} {93}},\ \bibinfo
  {pages} {011603(R)} (\bibinfo {year} {2016})}\BibitemShut {NoStop}%
\bibitem [{\citenamefont {Lee}\ and\ \citenamefont {Yang}(1957)}]{Lee58}%
  \BibitemOpen
  \bibfield  {author} {\bibinfo {author} {\bibfnamefont {T.~D.}\ \bibnamefont
  {Lee}}\ and\ \bibinfo {author} {\bibfnamefont {C.~N.}\ \bibnamefont {Yang}},\
  }\href {https://doi.org/10.1103/PhysRev.105.1119} {\bibfield  {journal}
  {\bibinfo  {journal} {Phys. Rev.}\ }\textbf {\bibinfo {volume} {105}},\
  \bibinfo {pages} {1119} (\bibinfo {year} {1957})}\BibitemShut {NoStop}%
\bibitem [{\citenamefont {Giorgini}\ \emph {et~al.}(2008)\citenamefont
  {Giorgini}, \citenamefont {Pitaevskii},\ and\ \citenamefont
  {Stringari}}]{Giorgini08}%
  \BibitemOpen
  \bibfield  {author} {\bibinfo {author} {\bibfnamefont {S.}~\bibnamefont
  {Giorgini}}, \bibinfo {author} {\bibfnamefont {L.~P.}\ \bibnamefont
  {Pitaevskii}},\ and\ \bibinfo {author} {\bibfnamefont {S.}~\bibnamefont
  {Stringari}},\ }\href {https://doi.org/10.1103/RevModPhys.80.1215} {\bibfield
   {journal} {\bibinfo  {journal} {Rev. Mod. Phys.}\ }\textbf {\bibinfo
  {volume} {80}},\ \bibinfo {pages} {1215} (\bibinfo {year}
  {2008})}\BibitemShut {NoStop}%
\bibitem [{\citenamefont {Haussmann}\ \emph {et~al.}(2007)\citenamefont
  {Haussmann}, \citenamefont {Rantner}, \citenamefont {Cerrito},\ and\
  \citenamefont {Zwerger}}]{Haussman07}%
  \BibitemOpen
  \bibfield  {author} {\bibinfo {author} {\bibfnamefont {R.}~\bibnamefont
  {Haussmann}}, \bibinfo {author} {\bibfnamefont {W.}~\bibnamefont {Rantner}},
  \bibinfo {author} {\bibfnamefont {S.}~\bibnamefont {Cerrito}},\ and\ \bibinfo
  {author} {\bibfnamefont {W.}~\bibnamefont {Zwerger}},\ }\href
  {https://doi.org/10.1103/PhysRevA.75.023610} {\bibfield  {journal} {\bibinfo
  {journal} {Phys. Rev. A}\ }\textbf {\bibinfo {volume} {75}},\ \bibinfo
  {pages} {023610} (\bibinfo {year} {2007})}\BibitemShut {NoStop}%
\bibitem [{\citenamefont {Regal}\ \emph {et~al.}(2004)\citenamefont {Regal},
  \citenamefont {Greiner},\ and\ \citenamefont {Jin}}]{Regal04}%
  \BibitemOpen
  \bibfield  {author} {\bibinfo {author} {\bibfnamefont {C.~A.}\ \bibnamefont
  {Regal}}, \bibinfo {author} {\bibfnamefont {M.}~\bibnamefont {Greiner}},\
  and\ \bibinfo {author} {\bibfnamefont {D.~S.}\ \bibnamefont {Jin}},\ }\href
  {https://doi.org/10.1103/PhysRevLett.92.040403} {\bibfield  {journal}
  {\bibinfo  {journal} {Phys. Rev. Lett.}\ }\textbf {\bibinfo {volume} {92}},\
  \bibinfo {pages} {040403} (\bibinfo {year} {2004})}\BibitemShut {NoStop}%
\bibitem [{\citenamefont {De~Rosi}\ and\ \citenamefont
  {Stringari}(2015)}]{DeRosi15}%
  \BibitemOpen
  \bibfield  {author} {\bibinfo {author} {\bibfnamefont {G.}~\bibnamefont
  {De~Rosi}}\ and\ \bibinfo {author} {\bibfnamefont {S.}~\bibnamefont
  {Stringari}},\ }\href {https://doi.org/10.1103/PhysRevA.92.053617} {\bibfield
   {journal} {\bibinfo  {journal} {Phys. Rev. A}\ }\textbf {\bibinfo {volume}
  {92}},\ \bibinfo {pages} {053617} (\bibinfo {year} {2015})}\BibitemShut
  {NoStop}%
\bibitem [{\citenamefont {Murthy}\ \emph {et~al.}(2014)\citenamefont {Murthy},
  \citenamefont {Kedar}, \citenamefont {Lompe}, \citenamefont {Neidig},
  \citenamefont {Ries}, \citenamefont {Wenz}, \citenamefont {Z\"urn},\ and\
  \citenamefont {Jochim}}]{Murthy14}%
  \BibitemOpen
  \bibfield  {author} {\bibinfo {author} {\bibfnamefont {P.~A.}\ \bibnamefont
  {Murthy}}, \bibinfo {author} {\bibfnamefont {D.}~\bibnamefont {Kedar}},
  \bibinfo {author} {\bibfnamefont {T.}~\bibnamefont {Lompe}}, \bibinfo
  {author} {\bibfnamefont {M.}~\bibnamefont {Neidig}}, \bibinfo {author}
  {\bibfnamefont {M.~G.}\ \bibnamefont {Ries}}, \bibinfo {author}
  {\bibfnamefont {A.~N.}\ \bibnamefont {Wenz}}, \bibinfo {author}
  {\bibfnamefont {G.}~\bibnamefont {Z\"urn}},\ and\ \bibinfo {author}
  {\bibfnamefont {S.}~\bibnamefont {Jochim}},\ }\href
  {https://doi.org/10.1103/PhysRevA.90.043611} {\bibfield  {journal} {\bibinfo
  {journal} {Phys. Rev. A}\ }\textbf {\bibinfo {volume} {90}},\ \bibinfo
  {pages} {043611} (\bibinfo {year} {2014})}\BibitemShut {NoStop}%
\bibitem [{\citenamefont {Henn}\ \emph {et~al.}(2009)\citenamefont {Henn},
  \citenamefont {Seman}, \citenamefont {Roati}, \citenamefont {Magalh\~aes},\
  and\ \citenamefont {Bagnato}}]{Henn09}%
  \BibitemOpen
  \bibfield  {author} {\bibinfo {author} {\bibfnamefont {E.~A.~L.}\
  \bibnamefont {Henn}}, \bibinfo {author} {\bibfnamefont {J.~A.}\ \bibnamefont
  {Seman}}, \bibinfo {author} {\bibfnamefont {G.}~\bibnamefont {Roati}},
  \bibinfo {author} {\bibfnamefont {K.~M.~F.}\ \bibnamefont {Magalh\~aes}},\
  and\ \bibinfo {author} {\bibfnamefont {V.~S.}\ \bibnamefont {Bagnato}},\
  }\href {https://doi.org/10.1103/PhysRevLett.103.045301} {\bibfield  {journal}
  {\bibinfo  {journal} {Phys. Rev. Lett.}\ }\textbf {\bibinfo {volume} {103}},\
  \bibinfo {pages} {045301} (\bibinfo {year} {2009})}\BibitemShut {NoStop}%
\bibitem [{\citenamefont {Ku}\ \emph {et~al.}(2012)\citenamefont {Ku},
  \citenamefont {Sommer}, \citenamefont {Cheuk},\ and\ \citenamefont
  {Zwierlein}}]{Ku12}%
  \BibitemOpen
  \bibfield  {author} {\bibinfo {author} {\bibfnamefont {M.~J.~H.}\
  \bibnamefont {Ku}}, \bibinfo {author} {\bibfnamefont {A.~T.}\ \bibnamefont
  {Sommer}}, \bibinfo {author} {\bibfnamefont {L.~W.}\ \bibnamefont {Cheuk}},\
  and\ \bibinfo {author} {\bibfnamefont {M.~W.}\ \bibnamefont {Zwierlein}},\
  }\href {https://doi.org/10.1126/science.1214987} {\bibfield  {journal}
  {\bibinfo  {journal} {Science}\ }\textbf {\bibinfo {volume} {335}},\ \bibinfo
  {pages} {563} (\bibinfo {year} {2012})}\BibitemShut {NoStop}%
\bibitem [{\citenamefont {Carr}\ \emph {et~al.}(2004)\citenamefont {Carr},
  \citenamefont {Shlyapnikov},\ and\ \citenamefont {Castin}}]{Carr04}%
  \BibitemOpen
  \bibfield  {author} {\bibinfo {author} {\bibfnamefont {L.~D.}\ \bibnamefont
  {Carr}}, \bibinfo {author} {\bibfnamefont {G.~V.}\ \bibnamefont
  {Shlyapnikov}},\ and\ \bibinfo {author} {\bibfnamefont {Y.}~\bibnamefont
  {Castin}},\ }\href {https://doi.org/10.1103/PhysRevLett.92.150404} {\bibfield
   {journal} {\bibinfo  {journal} {Phys. Rev. Lett.}\ }\textbf {\bibinfo
  {volume} {92}},\ \bibinfo {pages} {150404} (\bibinfo {year}
  {2004})}\BibitemShut {NoStop}%
\bibitem [{\citenamefont {Tan}(2008{\natexlab{c}})}]{Tan08a}%
  \BibitemOpen
  \bibfield  {author} {\bibinfo {author} {\bibfnamefont {S.}~\bibnamefont
  {Tan}},\ }\href@noop {} {\bibfield  {journal} {\bibinfo  {journal} {Annals of
  Physics}\ }\textbf {\bibinfo {volume} {323}},\ \bibinfo {pages} {2952}
  (\bibinfo {year} {2008}{\natexlab{c}})}\BibitemShut {NoStop}%
\bibitem [{\citenamefont {Braaten}(2012)}]{Braaten12}%
  \BibitemOpen
  \bibfield  {author} {\bibinfo {author} {\bibfnamefont {E.}~\bibnamefont
  {Braaten}},\ }\bibinfo {title} {Universal relations for fermions with large
  scattering length},\ in\ \href@noop {} {\emph {\bibinfo {booktitle} {The
  BCS-BEC Crossover and the Unitary Fermi Gas}}}\ (\bibinfo  {publisher}
  {Springer},\ \bibinfo {year} {2012})\ pp.\ \bibinfo {pages}
  {193--231}\BibitemShut {NoStop}%
\bibitem [{\citenamefont {Thomas}\ \emph {et~al.}(2005)\citenamefont {Thomas},
  \citenamefont {Kinast},\ and\ \citenamefont {Turlapov}}]{Thomas05}%
  \BibitemOpen
  \bibfield  {author} {\bibinfo {author} {\bibfnamefont {J.~E.}\ \bibnamefont
  {Thomas}}, \bibinfo {author} {\bibfnamefont {J.}~\bibnamefont {Kinast}},\
  and\ \bibinfo {author} {\bibfnamefont {A.}~\bibnamefont {Turlapov}},\ }\href
  {https://doi.org/10.1103/PhysRevLett.95.120402} {\bibfield  {journal}
  {\bibinfo  {journal} {Phys. Rev. Lett.}\ }\textbf {\bibinfo {volume} {95}},\
  \bibinfo {pages} {120402} (\bibinfo {year} {2005})}\BibitemShut {NoStop}%
\bibitem [{\citenamefont {Werner}\ \emph {et~al.}(2009)\citenamefont {Werner},
  \citenamefont {Tarruell},\ and\ \citenamefont {Castin}}]{Werner09}%
  \BibitemOpen
  \bibfield  {author} {\bibinfo {author} {\bibfnamefont {F.}~\bibnamefont
  {Werner}}, \bibinfo {author} {\bibfnamefont {L.}~\bibnamefont {Tarruell}},\
  and\ \bibinfo {author} {\bibfnamefont {Y.}~\bibnamefont {Castin}},\
  }\href@noop {} {\bibfield  {journal} {\bibinfo  {journal} {The European
  Physical Journal B}\ }\textbf {\bibinfo {volume} {68}},\ \bibinfo {pages}
  {401} (\bibinfo {year} {2009})}\BibitemShut {NoStop}%
\bibitem [{\citenamefont {Luciuk}\ \emph {et~al.}(2016)\citenamefont {Luciuk},
  \citenamefont {Trotzky}, \citenamefont {Smale}, \citenamefont {Yu},
  \citenamefont {Zhang},\ and\ \citenamefont {Thywissen}}]{Luciuk16}%
  \BibitemOpen
  \bibfield  {author} {\bibinfo {author} {\bibfnamefont {C.}~\bibnamefont
  {Luciuk}}, \bibinfo {author} {\bibfnamefont {S.}~\bibnamefont {Trotzky}},
  \bibinfo {author} {\bibfnamefont {S.}~\bibnamefont {Smale}}, \bibinfo
  {author} {\bibfnamefont {Z.}~\bibnamefont {Yu}}, \bibinfo {author}
  {\bibfnamefont {S.}~\bibnamefont {Zhang}},\ and\ \bibinfo {author}
  {\bibfnamefont {J.~H.}\ \bibnamefont {Thywissen}},\ }\href@noop {} {\bibfield
   {journal} {\bibinfo  {journal} {Nature Physics}\ }\textbf {\bibinfo {volume}
  {12}},\ \bibinfo {pages} {599} (\bibinfo {year} {2016})}\BibitemShut
  {NoStop}%
\bibitem [{\citenamefont {Qi}\ \emph {et~al.}(2021)\citenamefont {Qi},
  \citenamefont {Shi},\ and\ \citenamefont {Zhai}}]{Qi21}%
  \BibitemOpen
  \bibfield  {author} {\bibinfo {author} {\bibfnamefont {R.}~\bibnamefont
  {Qi}}, \bibinfo {author} {\bibfnamefont {Z.}~\bibnamefont {Shi}},\ and\
  \bibinfo {author} {\bibfnamefont {H.}~\bibnamefont {Zhai}},\ }\href
  {https://doi.org/10.1103/PhysRevLett.126.240401} {\bibfield  {journal}
  {\bibinfo  {journal} {Phys. Rev. Lett.}\ }\textbf {\bibinfo {volume} {126}},\
  \bibinfo {pages} {240401} (\bibinfo {year} {2021})}\BibitemShut {NoStop}%
\end{thebibliography}%

\end{document}